# The LCLS-II Photoinjector Laser Infrastructure

Hao Zhang[1,2,*], Sasha Gilevich[1], Alan Miahnahri[1], Shawn Alverson[1], Axel Brachmann[1], Joseph Duris[1], Paris Franz[1,3], Alan Fry[1], Jack Hirschman[1,3], Kirk Larsen[1], Randy Lemons[1], Siqi Li[1], Brittany Lu[2], Agostino Marinelli[1], Mikael Martinez[1], Justin May[1], Erel Milshtein[1], Krishna Murari[1], Nicole Neveu[1], Joseph Robinson[1], John Schmerge[1], Linshan Sun[2], Theodore Vecchione[1], Chengcheng Xu[1], Feng Zhou[1], and Sergio Carbajo [1,2,4,*]

[1]*SLAC National Accelerator Laboratory, Stanford University, Menlo Park, California 94025, USA*
[2]*Department of Electrical and Computer Engineering, University of California Los Angeles, 420 Westwood Plaza, Los Angeles, CA 90095, USA*
[3]*Department of Applied Physics, Stanford University, 348 Via Pueblo, Stanford University, CA 94305, USA*
[4]*Physics and Astronomy Department, University of California Los Angeles, 475 Portola Plaza, Los Angeles, CA 90095, USA*
*[*]haozh@g.ucla.edu; scarbajo@g.ucla.edu*

## Abstract

This paper presents a comprehensive technical overview of the Linac Coherent Light Source II (LCLS-II) photoinjector laser system, its first and foremost component. The LCLS-II photoinjector laser system serves as an upgrade to the original LCLS at SLAC National Accelerator Laboratory. This advanced laser system generates high-quality laser beams for LCLS-II, contributing to the instrument's unprecedented brightness, precision, and flexibility. Our discussion extends to the various subsystems that comprise the photoinjector, including the photocathode laser, laser heater, and beam transport systems. Lastly, we draw attention to the ongoing research and development infrastructure underway to enhance the functionality and efficiency of the LCLS-II, and similar X-ray free-electron laser facilities around the world, thereby contributing to the future of laser technology and its applications.

## I. INTRODUCTION

X-ray free-electron lasers (XFELs) have transformed X-ray science by providing ultrafast, coherent X-ray pulses with unprecedented peak and average brightness[1], enabling the study of fundamental dynamics in atomic, molecular, and optical (AMO) physics[2], condensed matter physics[3], biology[4] and chemistry[5]. In recent years, XFELs have emerged as a promising complement to third-generation synchrotron light sources, offering significantly higher brightness and shorter pulse durations than traditional X-ray sources[6,7]. The first XFEL, Free electron LASer (FLASH), was successfully operated in 2006 at Deutsches Elektronen-Synchrotron (DESY) in Hamburg[8], followed by the Linac Coherent Light Source (LCLS) in 2009, which produced 10keV X-rays and became the world's first hard X-ray FEL[9]. This generation of light sources delivers up to 10 orders of magnitude increase in brightness, spatial resolution into the nanometer scale, and femtosecond[10] temporal resolution. Japan also developed a compact XFEL (SPring-8 Angstrom Compact free-electron LAser, SACLA) emitting in the sub-ångström region in 2012[11]. However, low repetition rates (e.g. LCLS-I with 120 Hz, SACLA with 60 Hz) limit the potential output in ultrafast sciences[12–15]. Their relatively low repetition rates led to the development of the European XFEL[16–19], LCLS-II, and its future higher energy (-HE) extension, LCLS-II-HE. In response to the recommendation from the Department of Energy Office of Sciences' Basic Sciences Advisory Committee, LCLS-II is capable of operating at a 1 MHz repetition rate[13,20–23]. LCLS-II will reach an average brightness up to 4 orders of magnitude greater than LCLS and target X-ray production from ~200 eV to beyond 20 keV[13,14,21]. These extensive enhancements combined with the significant increase in repetition rate and new developments for attosecond pulse production[1] are ushering in a new era in studies of fundamental dynamics of energy in AMO physics[10,24–27], nanoscale dynamics[5,28–30], matter in extreme conditions[31–34], chemical sciences[35,36], and structural biology among many other ultrafast physical sciences[9,37–39], even inspiring endeavors such as single-particle imaging[40,41].

Typically, XFELs encompass a photoinjector, a linear accelerator, an undulator, and end stations, each component performing a vital role in the overall operation and output of the system[11,41,42]. The LCLS-II photoinjector is strictly limited to the components preceding the first linac, as shown in Figure 1a. This includes, but is not limited to, the photoinjector gun and the laser heater. the photoinjector is responsible for generating the initial electron distribution, which is characterized in a six-dimensional (6D) space. This encompasses three spatial dimensions (x, y, z) representing the electrons' positions, and three corresponding momentum dimensions, detailing their velocities or energy directions in each spatial axis. These electrons are then accelerated to higher energies using RF cavities and then stimulated by undulators to produce X-rays. As such, the photoinjector determines fundamental limits on the quality of the X-ray generation process[20,41,43,44]. Incorporating components such as a laser system, photocathode gun, RF injector, and bunch compressor, the LCLS-II photoinjector system facilitates the meticulous manipulation of the laser pulses that induce photoemission. This capability significantly enhances the electron beam quality and reduces pulse durations, thereby opening up new frontiers for cutting-edge research across various ultrafast X-ray scientific fields.[20,41].

This article offers a comprehensive overview of the LCLS-II photoinjector system and its operations, bridging any gaps left by fragmented literature. We will discuss potential engineering challenges and spotlight the R&D infrastructure, covering aspects like laser temporal and spatial shaping, visible-range responsive photocathode development, and adaptive spatial-temporal laser beam shaping. By charting the path for innovative photoinjector R&D, we underscore the photoinjector's pivotal role in the successful operation of LCLS-II and its forthcoming enhancements.

## II. LCLS-II PHOTOINJECTOR SYSTEMS

### IIa. Overview

The LCLS-II photoinjector system represents a significant improvement over the LCLS-I system, as shown in Figure 1a, offering greater output photon energy (from 0.25 to 5 keV) and photon efficiency (produce >$10^{10}$ photons per pulse up to ~5 keV photon energy)[20,41]. The photocathode laser system generates ultraviolet (UV) pulses illuminating a photocathode inside a high-field RF gun. These UV pulses generate photoelectrons that are then accelerated by the RF field to produce high average power, efficient, and high-quality electron bunches. The UV pulses influence both the duration and emittance of the electron bunch[9,14]. Additionally, the UV production system incorporates advanced diagnostics and controls, enabling users to precisely tune the beam parameters to meet a broad range of operational requirements. Specifically, the system employs pre-upconversion programmable spectral amplitude and phase IR shaping using both acousto-optic programmable dispersive filters and spatial light modulators. These devices allow for high-rate control, and, when linked to the application-appropriate diagnostic, can form a high-speed feedback system for tailoring the upconverted UV beam. In the following sections, we will delve into the operation of various laser systems, including the photocathode laser system, the laser heater laser system, and the laser-electron timing and synchronization system. We also address the engineering challenges of the laser system.

### IIb. Photocathode Laser System

The diagram of the current LCLS-II photocathode drive laser system is shown in Figure 1a (red dashed box) and Figure 1b. The infrared (IR) front-end system is a Ytterbium-based chirped pulse amplifier (CPA) system capable of producing up to 50 µJ per pulse at 1030 nm and repetition rates up to 1 MHz at a transform-limited duration of 330 fs[41]. Additionally, the high-energy IR pulse duration can be adjusted continuously between approximately 330 fs and 30ps. The mode-locked oscillator seeding the CPA is phase-locked to the facility RF system (shown in Figure 1b -RF locked timing system, more detail provided in IId section) backing the timing of all components along the accelerator. To allow for temporal pulse shaping diagnostics, such as an optical cross-correlator, the laser system is complemented by a short pulse module, shown in Figure 1b. This module takes a signal from the oscillator (~80 fs pulse duration, 12 nJ/pulse energy) and amplifies it through a single-pass fiber amplification, thus producing short IR pulses with a duration below 75 fs at the repetition rate of the oscillator (46.5 MHz). This unit ensures that the laser pulses maintain the intended characteristics (~50 fs FWHM, 50 µJ per pulse, center at 1030 nm), verifying the system's optimal operation. The laser system will include a programmable spectral phase and amplitude shaper allowing IR spectrum manipulation and assisting in the temporal pulse shaping of the UV pulses (more details in Section IVc). The main required target parameters for generating the high-quality UV beam on the cathode are presented in Table I. These parameters are determined through rigorous tests and simulations to achieve optimum quantum efficiency of the electrons from the cathode, as discussed in the previous works[45,46]. LCLS operates 24×7, running for weeks to months without much intervention. Optical component reliability varies with repetition rates. Tests were performed during the commissioning and early operation phases of LCLS-II, thereby limiting the repetition rate and temporal format to varied rates from 1Hz to 100kHz, with short bursts up to 1MHz. After weeks, minor pointing shifts appear 20 meters downstream but are adjusted remotely using upstream mirrors. Some components, like the UV grating and waveplate, require replacement based on operation time and repetition rates. The energy jitter at the gun table is 4% after ~40 meters from the UV crystal.

Table I. Laser beam requirements on the cathode

| Parameter | Nom. value | Min | Max | Unit |
| --- | --- | --- | --- | --- |
| Operating Wavelength | 257.5 | - | - | nm |
| Pulse repetition rate | 0.625 | 0 | 0.929 | MHz |
| UV pulse energy at the cathode | 0.1 | - | 0.3 | µJ |
| UV beam power at the cathode (at 1 MHz) | 0.1 | - | 0.3 | W |
| Beam size on the cathode (FWHM) | 0.8 | 0.2 | 2 | mm |
| Pulse Duration | 30 | 20 | 60 | ps |
| Temporal Shape | Gaussian or Flat-top with maximum 2 ps rise/fall time | | | |
| Spatial Shape | Apodized Gaussian or flat-top | | | |

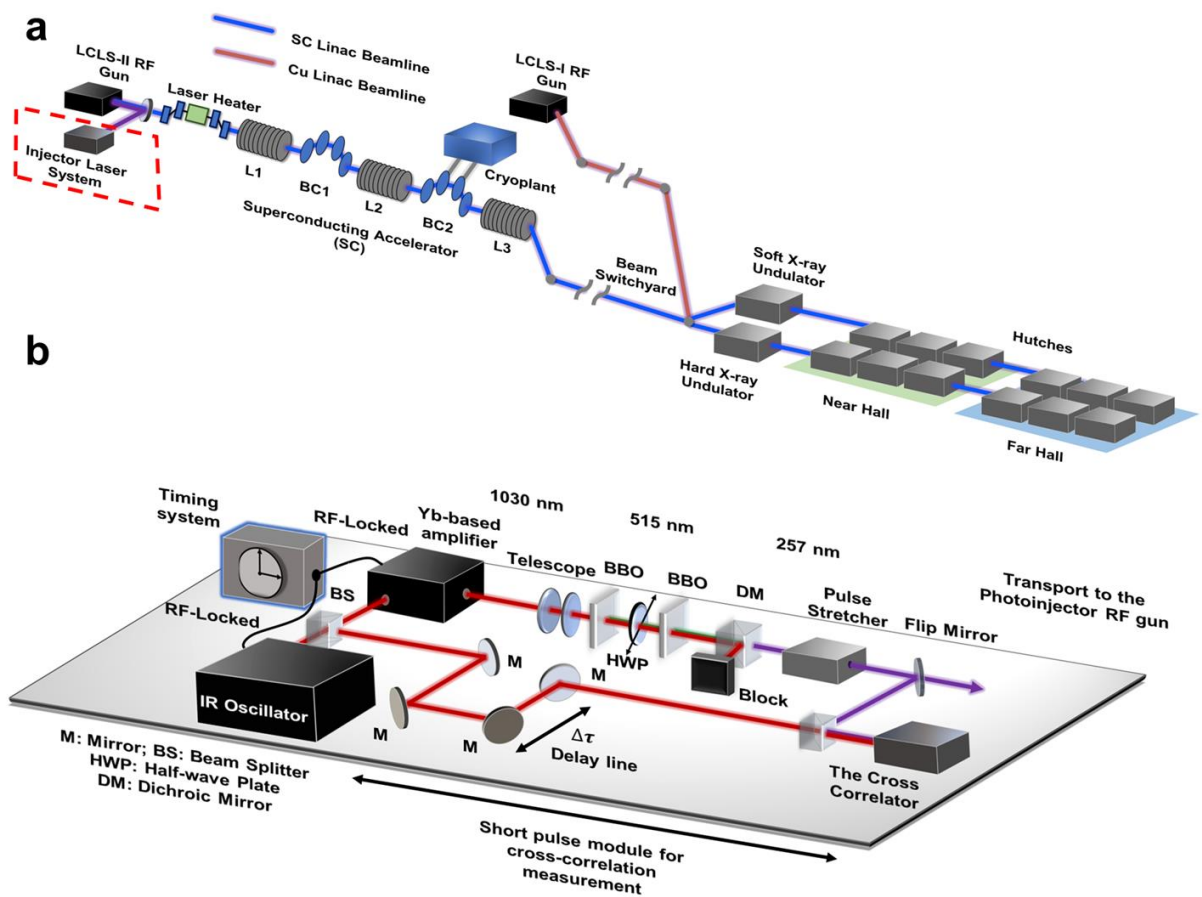

*Figure 1. (a) Streamlined diagram of the complete LCLS-II setup, extending from the photoinjector to the near/far experiment halls, not depicted to scale. 'L': linac, 'BC': bunch compressor. The injector laser system is in sector 0. (b) Simplified diagram of the photocathode drive laser system. The laser, UV conversion unit, energy attenuator, and conditioning system that adjusts pulse size and duration are located in the laser room in the housing upstream of the accelerator. Diagnostics include power meters and cameras located in the laser room and on the gun tables and cross-correlator in the laser room. The short pulse module is*

To convert the IR laser pulses to UV for photoemission, two critically phase-matched barium borate (BBO) crystals are used, as shown in Figure 1b. The first is a 3 mm long second harmonic generation (SHG) crystal that generates 515 nm pulses through frequency doubling. The second BBO is a 1 mm long SHG crystal that produces the required 257.5 nm pulses for photoemission. The LCLS-II photoinjector uses CsTe as its photocathode material, which offers high quantum efficiency and the ability to work at the extremely high accelerating gradients used in the photoinjector system[47]. The second SHG crystal is enclosed in a temperature-regulated environment to avoid varying thermal lensing effects at different repetition rates. For near-transform limited IR pulse SHG, the conversion efficiency is achieved between 50% to 65%, while the fourth harmonic generation (FHG) conversion efficiency from IR to UV is approximately 8% to 20%. These variations are attributed to different IR repetition rates and the resulting thermal load on both the first and second SHG crystals.

Although compressed IR pulses are effective for achieving high conversion efficiency and short UV pulse durations, the desired beam parameters for the photocathode necessitate UV pulses with longer durations, in the realm of several picoseconds (as detailed in Table I). When considering the target charge levels (100 - 500 pC) and the transverse dimension of the electron beam (refer to Table II), a more nuanced understanding of the electron-beam generated at the photocathode material is required. This includes not only the impact of these longer UV pulses on the photoemission process but also a detailed analysis of space charge effects. Specifically, the space charge limits can significantly influence the initial electron beam dynamics, such as emittance growth and beam density distribution. These factors are crucial for optimizing the photocathode performance and require careful consideration in the context of the LCLS-II photoinjector system. This affects the interaction time of each photon with the photocathode material and its overall efficiency. This prolonged interaction time with the photocathode, facilitated by the extended UV pulse durations, leads to a reduced energy spread in the electron beam and lower emittance. Simultaneously, it enhances the temporal resolution of the X-ray pulses generated by the system. It also augments the overall stability and reproducibility of the resulting electron beam. In typical collinear SHG and FHG processes using chirped IR beams, there's a tendency to induce temporal intensity modulations in the UV beam, often leading to reduced conversion efficiencies. To overcome this, our baseline approach employs FHG with fully compressed IR beams, combined with the expansion of UV bandwidth using a Treacy-type 4-pass reflective grating stretcher. This setup allows us to adjust the path length of the stretcher, enabling pulse duration variation from 5 to 30 picoseconds in our current configuration. Notably, the grating we use achieves a single-pass first-order efficiency of approximately 75%, which translates to about 30% transmission through the stretcher. Uniquely, our approach diverges from the conventional compression role of the Treacy arrangement. We utilize it for pulse stretching, as our cathode is effectively agnostic to the chirp sign, whether positive or negative. Looking ahead, enhancing the stretcher's efficiency could involve transitioning to transmission gratings, which offer an estimated 80% efficiency. Importantly, while considering this upgrade, the pulse energy in our setup remains below the threshold that would cause substrate degradation under 257.5 nm radiation, ensuring the longevity and integrity of the grating. However, under the high power environment, the durability of the substrate becomes a critical concern.

## IIc. Laser Heater Laser System in LCLS-II

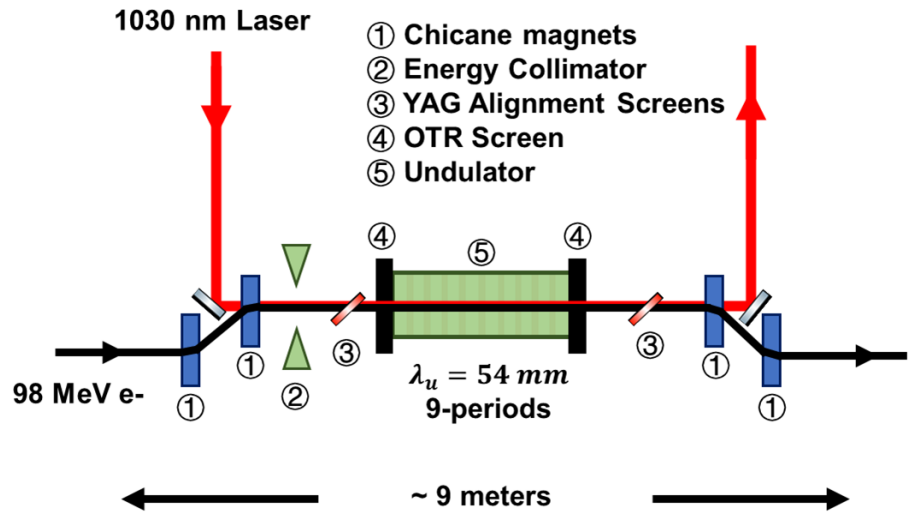

*Figure 2. The diagram illustrates the configuration of the LCLS-II laser-heater system, which includes the 1030 nm laser, chicane magnets, optical transition radiation (OTR) screen, the energy collimator, pop-in YAG alignment screens (to align the laser and the electron beam), and the undulators[48,49].*

The laser heater system is located just downstream of the L0 accelerator section at roughly 100 MeV. The magnetic bunch compressors used to generate the bright electron beam required for XFELs, are prone to microbunching instability (MBI). MBI occurs when small density fluctuations within the electron bunch are amplified as the bunch is compressed, leading to the formation of substructures or 'microbunches' within the main electron bunch. This process can significantly increase the slice energy spread – the variation in energy of the electrons within a specific longitudinal slice of the bunch. When the slice energy spread exceeds the maximum acceptable limit, it adversely affects the quality of the X-ray beam produced by the XFEL.The increased energy spread can lead to reduced peak brightness and coherence of the X-ray beam, ultimately impacting the performance and the range of scientific applications achievable with the XFEL[48–54]. To mitigate this instability, a laser heater (LH)[53,54] is used to introduce a small amount of energy spread to the electron beam. The role of the laser heater is to add a minimal, controlled level of energy spread to the electron beam. This introduction of energy spread is a deliberate technique to achieve Landau damping of the microbunching instability. The concept may initially appear paradoxical, given that microbunching instability itself leads to an unwanted increase in energy spread, which is detrimental to the electron beam's quality. However, the key distinction lies in the nature and purpose of the energy spread introduced by the laser heater. Unlike the uncontrolled and destabilizing energy spread caused by microbunching, the laser heater's contribution is finely tuned and serves a preventive purpose. By adding this small amount of energy spread early on, it damps the growth of microbunching instability, thus preserving the beam's high brightness. The controlled energy spread acts as a stabilizing factor, preventing the microbunching instability from escalating to a level where it could significantly disrupt the electron beam, thereby maintaining the optimal performance of the XFEL. Such a system has been implemented at LCLS[54], a common standard feature in all short-wavelength FEL projects.

While laser heaters have historically been utilized in XFELs for microbunching suppression, their role is rapidly evolving with advancements in electron beam shaping mechanisms[55], notably the recent demonstration of attosecond bunching at LCLS. The LCLS-II, with its superconducting accelerator and heightened average power capacities for both electron and laser beams, necessitates cutting-edge adaptations in laser technology, pushing beyond the confines of established state-of-the-art solutions (exhibits an average brightness that is 10,000 times greater than that of LCLS-I and operates at a repetition rate 8,000 times faster, delivering up to one million pulses per second)[32]. The current LCLS-II photoinjector system also incorporates the LH system to generate an uncorrelated energy spread in the electron beam[55], as shown in Figure 2. The upgraded LH design utilizes a similar undulator as in LCLS. The photoinjector system is equipped with two lasers, one used as the drive laser as shown in Figure 1b, and the other one used for the LH, as shown in Figure 2. The selection of a 1030 nm wavelength for the LH laser is primarily based on the accessibility of high-power, high-repetition-rate sources such as Yb-doped glass fibers. The energy of the electron beam is determined by the acceleration from the first cryomodule. Primarily, the energy of the electron beam and the wavelength of the laser dictate the specifics of the laser heater undulator, which is an existing undulator with a 54 mm period and a moderate number of undulator periods ($N_u = 9$). This arrangement is intended to broaden the bandwidth of the undulator resonance condition by a few percent. The adjustable-gap heater undulator is capable of functioning at the higher spectrum of the injector electron beam's energy range, specifically at 120 MeV. This is achieved by narrowing the undulator gap to its minimum (3.2 cm), consequently amplifying the peak magnetic field to 0.3 T. Diagnostic tools (YAG alignment screens) for both the laser and the electron beam are positioned on either side of the undulator, enabling alignment of the laser and electron beam. Traditionally, the transverse shape of the laser beam used in LH at current LCLS-I is Gaussian. We investigated the application of a Laguerre-Gaussian 01 (LG01) mode laser, characterized by its distinctive donut-shaped beam profile, in a LH and assessed its effectiveness in suppressing MBI as well as its influence on FEL performance[56,57]. The laser beam, initially characterized by a Gaussian transverse profile, was transformed into an LG01 mode through the use of a spiral phase plate (SPP)[56]. This SPP introduces a helical phase pattern to the laser beam, incrementally modifying the beam's phase to achieve a cumulative change of $2\pi$. As a result of this phase adjustment, the beam's central field amplitude becomes nullified. We found that the LG01 mode LH induced a Gaussian-shaped energy distribution of the electron beam, which improved the suppression of the final microbunching gain. This is critical for the future development of clean, high spectral-brightness FEL pulses.

The effectiveness of the LH in increasing the energy spread of the electron beam is constrained by two key factors: firstly, the maximum permissible root mean square (rms) energy spread in the FEL's undulator section, where the electron beam interacts with the magnetic field to produce X-rays; and secondly, the total magnetic compression factor. The magnetic compression factor refers to the degree by which the electron beam is compacted in the FEL's magnetic bunch compressors. These compressors use magnetic fields to decrease the longitudinal spread of the electron bunch, thereby increasing its current. This factor is crucial because it influences how much the LH can safely increase the energy spread without adversely affecting the beam's coherence and the quality of the X-ray output. For the baseline beam with a charge of $Q = 100$ pC, the LH-induced heating should be kept below keV rms to effectively suppress the microbunching instability. Table II provides the key parameters for the LCLS-II Laser Heater system[55,57,58], which provides a detailed overview of the LH parameters, specifically tailored for handling electron bunches with a charge of 100 pC at a repetition rate of 1 MHz. The table also outlines specifics about the undulator, including its gap, period, and peak magnetic field. Laser characteristics, such as wavelength, pulse length, and power specifications in terms of peak and average, are also enumerated. Various parameters have specified ranges, indicating the system's flexibility and operational limits. By adjusting the magnetic undulator gap from 4.3 cm to 3.25 cm, we can maintain the undulator resonant condition accommodating the varying injector energy from 90 MeV to 120 MeV. Figure 3 demonstrates the necessary laser pulse energy in relation to the total energy of the electron beam, designed to achieve a 6 keV rms energy spread across the beam.

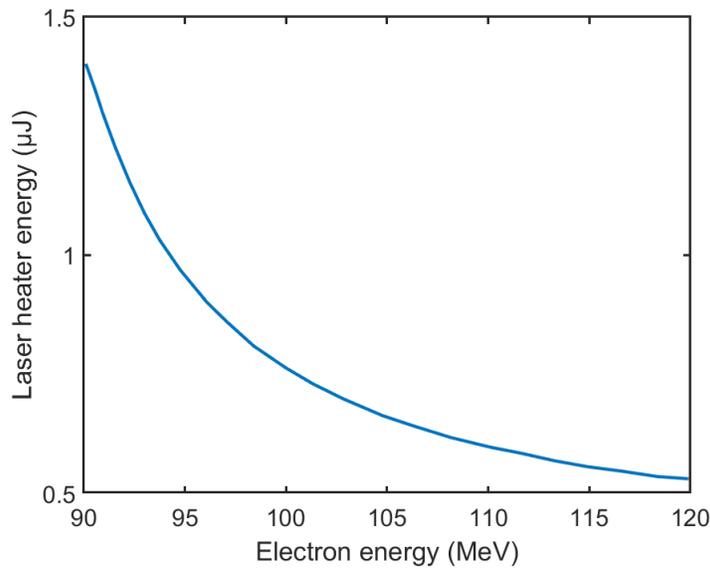

*Figure 3. The calculated relationship between laser pulse energy and electron beam energy, specifically showcasing a 6 keV rms energy spread induced by the LH system in the case of a 100 pC bunch.*

Table II. Key parameters for the LCLS-II Laser Heater system

| Parameter | Nominal | Range | Unit |
| --- | --- | --- | --- |
| Electron beam energy | 98 | 90-120 | MeV |
| Betatron functions (at LH undulator center) | 10 | 8-12 | m |

| Parameter | Nominal | Range | Unit |
| --- | --- | --- | --- |
| Normalized transverse emittance (used in these LH calculations) | 0.3 | 0.1-0.7 | μm |
| Electron beam transverse rms sizes (at LH undulator center) | 130 | 80-200 | μm |
| Chicane dipole bend angles | 0.022 | - | rad |
| Chicane dipoles lengths | 0.124 | - | m |
| Drift from 1st-to-2nd and 3rd-to-4th dipole | 3.28 | - | m |
| Dispersion (at undulator) | 7.5 | - | cm |
| Horizontal offset of undulator from linac axis | 7.5 | - | cm |
| Momentum compaction (over full chicane) | 3.5 | - | mm |
| Undulator gap (minimum 3.0 cm) | 4.1 | 4.3-3.20 | cm |

| Parameter | Nominal | Range | Unit |
|---|---|---|---|
| Undulator period | 5.4 | - | cm |
| Undulator parameter | 0.9 | 0.8-1.49 | - |
| Undulator peak magnetic field | 0.18 | 0.16-0.30 | T |
| Number of undulator periods | 9 | - | - |
| Laser wavelength | 1030 | - | nm |
| Laser beam diameter (middle of undulator) | 195 | 120-300 | μm |
| Rayleigh length | 46 | 18-110 | cm |
| Laser pulse duration | 20 | 10-30 | ps |
| Beam rms energy spread induced by Laser Heater | 6 | 0-20 | keV |
| Laser pulse energy at undulator | 1 | 0-15 | μJ |
| Laser pulse peak power | 0.05 | 0-1.5 | MW |

| Parameter | Nominal | Range | Unit |
|---|---|---|---|
| at undulator | | | |
| Laser average power at undulator (at 1 MHz) | 1 | 0-15 | W |

## IId. Timing and Synchronization System

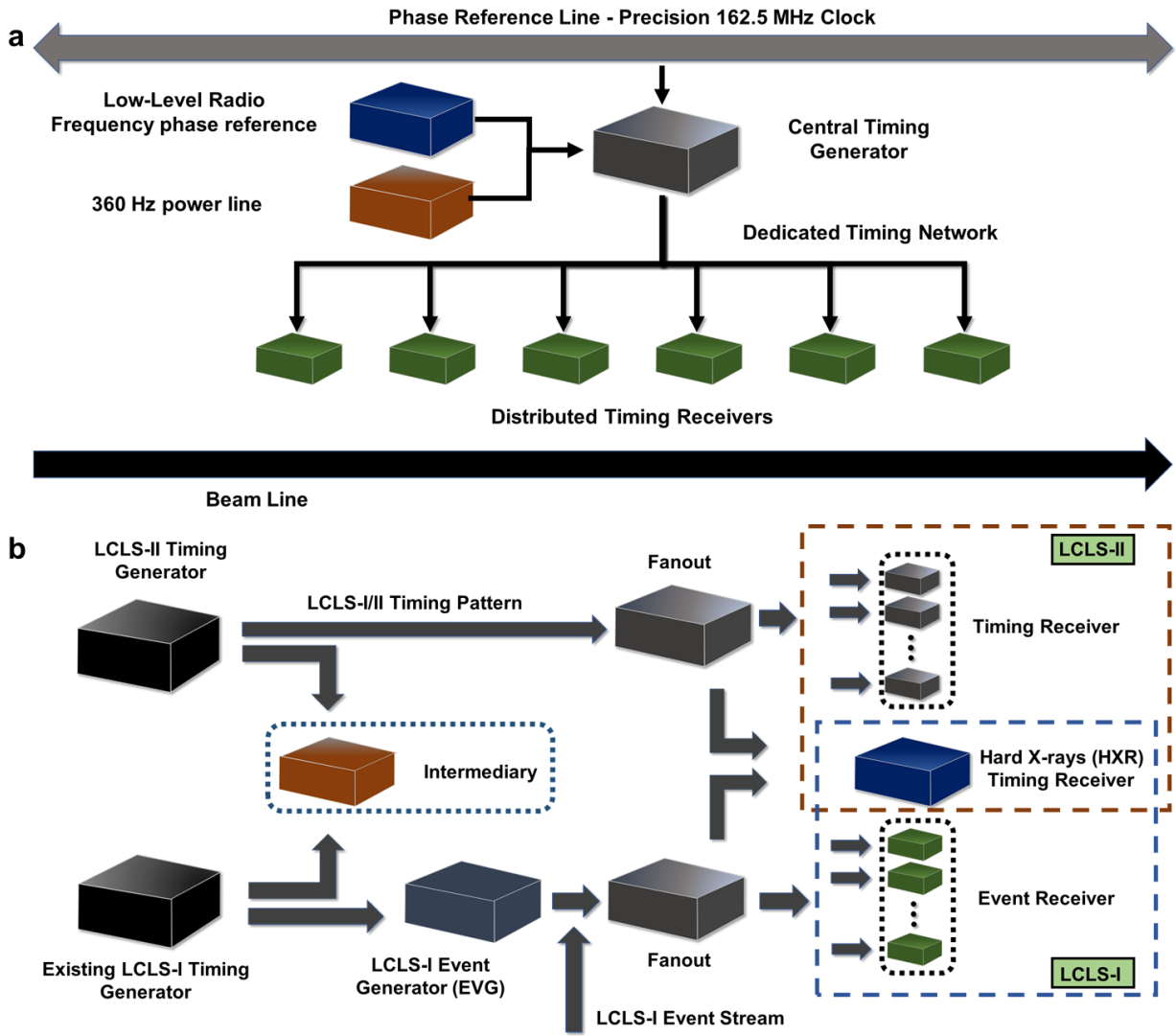

*Figure 4 (a). Generic layout for a central timing generator distributing timing information to receivers distributed along the beam line. (b) Interoperability of the LCLS and -II timing systems (shown in Figure 1).*

The optimal performance of the photoinjector laser system at the LCLS and -II depends on precise synchronization between them to ensure the electron bunches are accelerated at the right RF phase and arrive at the undulator at the correct time. To achieve this, the system employs the laser timing system (LTS), a specialized timing module that provides precise timing RF signals for the laser system. Figure 4a shows the general layout for a central timing generator distributing timing information to receivers distributed along the beam line. The core operation of the entire facility is synchronized to a master oscillator (MO) functioning at 162.5 MHz. To maintain precise timing, the 8th harmonic of the MO's frequency is extracted and disseminated through the phase reference line (PRL) system, which is integral to the laser timing system. Furthermore, the specific harmonics that dictate the beam's repetition rate are tailored to align with the requirements of the LLRF system, a crucial component of the primary superconducting linear accelerator. To increase overall capacity and enhance the capabilities of the LCLS facility, it is essential that the LCLS can operate in a standalone mode, independent of LCLS-II. This allows for greater flexibility in the operation of the light source. Consequently, the LCLS timing system should be designed as dependent on the main timing pattern from the LCLS-II, as depicted in Figure 4b. The laser timing system is also integrated with the event timing system, a crucial component that functions using an Event Generator (EVG). This EVG translates the timing pattern into specific events, which are then serially broadcast to distributed event receivers throughout the facility. These events originate from the central timing pattern, which is synchronized with the MO. The purpose of these events, often referred to as 'coarse timing triggers,' is to provide initial timing signals to various parts of the facility. These triggers ensure that the operations across different systems are synchronized with the overall timing pattern set by the MO. This synchronization is vital for maintaining the coordinated functioning of the entire facility, particularly in operations that require precise timing alignment, such as the synchronization of the laser system with other components of the accelerator. The LCLS-II system employs a phase reference line designated to distribute the 1300 MHz and 3900 MHz low-power-level radio frequency (LLRF) references. These frequencies are specifically designated for the operation of longitudinal phase-space linearizer (3ω linearizer) and deflectors within the system. The LLRF references are essential for maintaining precise control and synchronization of the radio frequencies used in these components, ensuring optimal performance of the LCLS-II system. The stringent timing requirements associated with these signals have been outlined as follows: The timing jitter, integrated over a frequency range from 50 Hz to 5 kHz, is specified at 10 fs. Short-term stability of the timing signals, observed over a duration of 1 second, mandates a precision of 1 fs. Over an extended period of 1 day, the stability requirement is defined at 1 ps. It is important to note that the UV beam is directly generated from the primary laser system. Consequently, these timing jitter specifications for the laser system also apply to the UV beam.

The laser timing system comprises two primary functions: first, phase-locking the laser to the machine's reference signal, ensuring synchronization with the accelerator's timing. Second, precisely and repeatably adjusting the timing of the laser's firing sequence. This adjustment, referred to as 'moving the laser time,' involves shifting the laser's pulse emission to specific, predetermined moments. This precise timing control is essential for the laser to interact effectively with the electron beam at the exact intervals required for optimal operation of the system. To phase-lock the laser, the fundamental repetition rate of the laser is extracted using a photodiode detector and measured with a Rubidium reference disciplined frequency counter. The fundamental repetition rate of the laser is picked to be 1/20th of the PRL frequency (65 MHz). A feedback loop is used to tune the repetition frequency to less than 1 Hz off from the desired value. Then the 40th harmonic of the laser fundamental frequency (~37.14 MHz) is extracted and locked to the phase reference line signal using an FPGA-based data acquisition module. The laser timing system measures the relative time between the laser pulse after the amplifier against a reference trigger from the event timing system by using a time interval counter. The timing can be varied using a combination of the event timing signal that triggers the laser amplifiers and the steady state error of the phase-locked loop. The value from the time interval counter is used to determine if the correct amount has been shifted. Both photocathode and LH lasers are locked using the same method described above. They are synchronized together by locking to the phase reference line signal and can achieve time overlap by shifting the timing of individual lasers.

A feedback loop continuously monitors the timing of the laser system using data from the bunch arrival time monitor (ATM) and other sensors, making adjustments as necessary to maintain synchronization with the electron bunches. The LTS provides a reliable and highly precise timing solution, enabling the photoinjector laser system to operate with sub-picosecond-level timing accuracy and ensuring the electron bunches arrive at the undulator at the correct time.

## IIe. Engineering Challenges in Photocathode Laser System

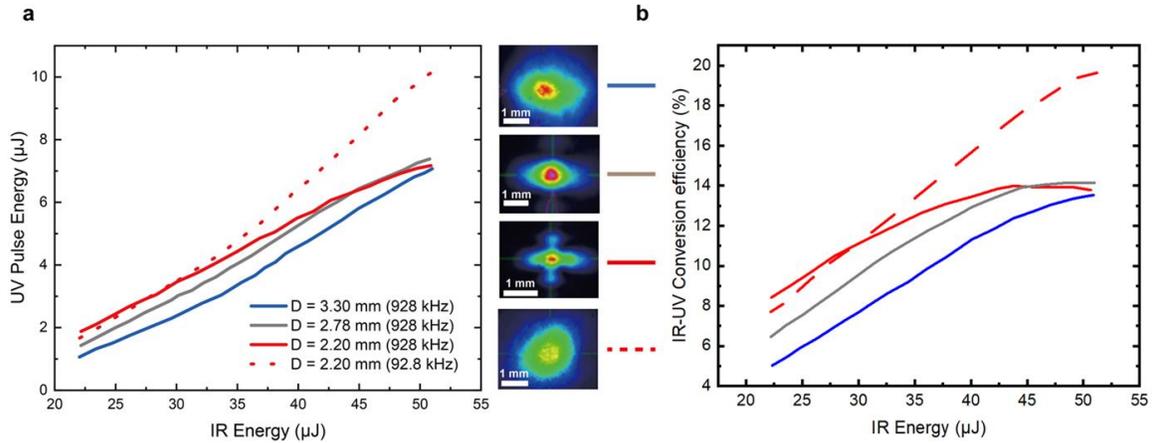

*Figure 5. (a) UV pulse energy and (b) IR-UV conversion efficiency by optimizing the SHG beam size in the the second SHG crystal. High repetition rates require finding a compromise between thermal stability, adequate spatial shape, and conversion efficiency. Inset images show the transverse beam shapes from diameter 3.30 mm to 2.20 mm. IR-UV conversion efficiency across different SHG beam diameters in the second SHG crystal at 928 kHz, with comparative data at 92.8 kHz for UV pulse energy and efficiency.*

The combination of Watt-level UV power operation, substantial UV absorption, and low thermal conductivity in the BBO crystals creates significant challenges for thermal management in the second SHG crystal. Figure 5a presents the UV pulse energy for three distinct SHG beam diameters at a high repetition rate of 928 kHz. At power levels below approximately 5.5 to 6 W, a reduction in beam size leads to an increase in UV pulse energy. However, at higher UV power levels, where thermal effects start impacting the FHG process, all three beam diameters converge to produce similar pulse energy outputs. In Figure 5b, at repetition rates under 100 kHz and with SHG beam diameters smaller than 2.8 mm, the UV conversion efficiency is notably higher than at rates above 100 kHz. With larger SHG (and consequently UV) beam sizes, where thermal effects are minimized due to improved heat extraction, the UV conversion efficiency remains relatively consistent. Additionally, at these lower repetition rates, optimizing the conversion efficiency is achievable through the reduction of beam size. At repetition rates above 100 kHz, we observed unstable UV power, poor beam quality characterized by deviations from the ideal Gaussian profile, and pronounced thermal lensing effects. To mitigate these issues, we optimized the SHG beam size within the the second SHG crystal for each specific power level through the implementation of a zoom telescope. By adjusting the SHG beam diameter from 1.5 mm (for repetition rates below 10 kHz) to 2.8 mm (for 1 MHz), we achieved stable power outputs up to approximately 6 W at 1

MHz. Furthermore, we quantified the improvement in beam quality by measuring the spatial profile's deviation from an ideal Gaussian shape, where we noted a significant enhancement across all required repetition rates. The detailed results demonstrating these enhancements in power stability and Gaussian pulse spatial quality are illustrated in Figure 5. While the figure also shows results for a 3.3 mm beam diameter, this diameter was part of a wider exploration and not central to our discussion on stable power levels. Our focus on the 1.5 mm to 2.8 mm range was driven by the need to balance power stability and beam quality across the required repetition rates. In scenarios where the repetition rates exceeded 300 kHz, we encountered a trade-off among thermal stability, adequate spatial shape, and conversion efficiency. As a result, in these high repetition rate conditions, the conversion efficiency experienced a reduction of approximately 20% compared to our maximum efficiency, which was achieved at lower repetition rates. However, even with this reduced conversion efficiency, we were able to meet the final power specification on the cathode (see Table I, ~0.1 W). For the future improvement of the conversion efficiency, we are considering various nonlinear conversion approaches, including noncollinear sum frequency generation[59–61] and four-wave mixing in a gas-filled hollow-core fiber[62].

### III. Beam Transport in LCLS-II

Since the RF gun has been installed in a radiation-shielded area[41], which is adjacent to the laser room, we require a stable laser transport system to deliver the generated UV beam generated from the laser system efficiently and securely through evacuated transport tubes (about 20 meters), as shown in Figure 6.

### IIIa. UV Beam Transport

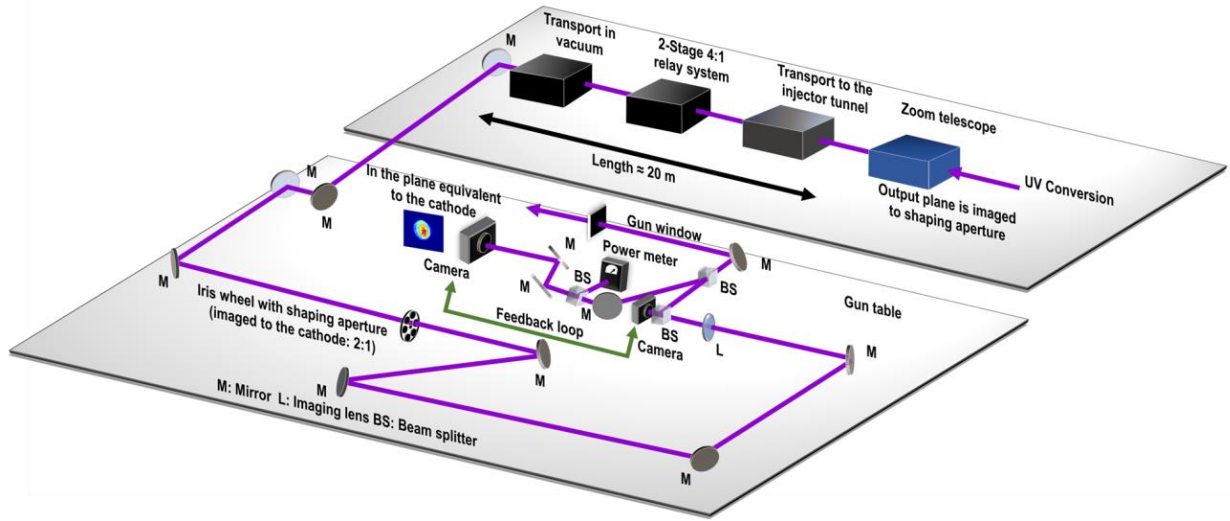

*Figure 6. The laser beam transport system for LCLS-II. (inset): The laser beam profile is monitored by the virtual cathode camera.*

The beam transport system adopts a three-lens zoom telescope, which facilitates adjustments to the beam size at the aperture. Then the beam goes pass a beam attenuator, comprised of a half-wave plate and a polarizer, situated downstream of the zoom telescope. A two-stage relay system with a demagnification ratio of 4:1 is used to image the output plane of the laser system (positioned downstream of the zoom telescope) onto a transverse beam-shaping aperture situated on the RF gun table. This relay system comprises turning mirrors and relay lenses, all housed within evacuated boxes. These evacuated boxes are combined with the transport tubes to create a unified vacuum system, which ensures the stability and integrity of the laser beam path.

In the subsequent stage, the transverse beam shaping aperture is imaged onto the cathode via the iris wheel with shaping aperture, achieving a 2:1 demagnification. This process intentionally overfills the aperture to ensure the creation of an apodized Gaussian shape, or in certain instances, a super-Gaussian/flattop profile on the cathode surface. In the transport system, we adopt the "virtual cathode" concept[63]. This needs the placement of a camera, replicating the cathode's position, to continuously monitor the beam's spatial shape and position with respect to the cathode. By working in conjunction with the virtual cathode and an additional camera, as shown in Figure 6, this feedback loop can ensure consistent positioning and direction of the laser beam inside the gun. In Table IV, we provide a summary of the transmission efficiency across various segments of the transport system. Before the beam transport, the UV energy measures around 1.8 µJ, and after transport, it is approximately 1.4 µJ. The transmission efficiency of the transport system is approximately 80%, likely constrained by fluorescence in the UV-grade fused silica transport optics. The minimum UV power stability is 1% and the minimum point stability is 10 µm. To improve this efficiency, we plan to use higher-quality optical materials[64], as we will introduce in the IIIb subsection. Additionally, we will consider enlarging the beam size on the transport optics to further enhance transmission.

Table IV. Transmission through the transport

| Device | Transmission |
| --- | --- |
| Beam Splitters for Diagnostics | 85% |
| Attenuator | 92% |
| Stretcher | 40% |
| Transport to injector tunnel | 80% |
| Transverse shaping aperture | 30% |

**IIIb. Degeneration Test**

To guarantee dependable functionality of high-power UV beams for continuous, 24/7 operation in future facilities, we have embarked on extensive long-term evaluations of various UV optical materials. These tests are specifically designed to assess the materials' performance and resilience under high-intensity UV conditions, characterized by peak powers ranging approximately from 0.4 to 2.5 watts and a beam diameter of about 1.2 mm. This analysis aims to establish a foundation for the reliable and sustainable use of these materials in advanced, high-demand UV applications. Regular UV-grade fused silica, which is commonly used by manufacturers, did not show significant transmission degradation during our tests (which lasted tens of hours). However, we observed significant degradation in the spatial beam quality due to the formation of color centers when the fluence was 120W/cm$^2$. We observed degradation in spatial shape after 18 hours of continuous illumination at this fluence and after 190 hours at a lower fluence of 25W/cm$^2$. These results are currently not operationally acceptable. We are now conducting similar experiments using higher purity non-crystalline silica glasses, such as Corning 8655, and calcium fluoride, such as Corning LDG CaF$_2$. We expect to achieve long-term high throughput operation without color center formation with Corning 8655, which has a purity of <1ppm of OH content. In the future, we will consider MgF2 and LiF as well[64]. The summarized optics damage tests are shown in Table III.

Table III. The results of optics damage tests

| Test Results (peak power ~0.4-2.5W) | UV-grade regular fused silica OH content 800 – 1000 ppm | Corning fused silica 8655. OH content <1 ppm | Corning Laser Durable CaF$_2$ |
|---|---|---|---|
| Destruction time at ~120 W/cm$^2$ | < 18 hours | Surface damage at 24 hours | No damage at 24 hours |
| Destruction time at ~80 W/cm$^2$ | ~24 hours | ~75 hours | >216 hours |
| Destruction time at ~60 W/cm$^2$ | <72 hours | ~116 hours | >116 hours |

We used the Carbide laser's uncompressed output (~50 fs FWHM, 50 µJ per pulse, center at 1030 nm) to run the window damage test. The 1030 nm IR laser from Carbide goes through a frequency conversion via nonlinear crystals, resulting in a UV laser centered at 257 nm wavelength.

## IV. Laser-FEL R&D Infrastructure in LCLS-II Photoinjector

### Overview

Research and development (R&D) lasers are used to study and improve the performance of lasers and their applications. LCLS R&D lasers are used to develop new LCLS laser technologies, improve existing laser systems, and explore new potential applications of FEL laser technology. This chapter discusses the R&D infrastructure for the photoinjector laser used in the LCLS-II facility. We focus on tailoring the laser pulse to optimize the quality and brightness of the electron beam, minimize transverse emittance growth, and also discuss the development of visible-range responsive photocathodes. These photocathodes are being improved to enhance their quantum efficiency – the effectiveness with which they convert incident laser light into electron beams. The R&D projects are still under testing and will be integrated to the LCLS-II in the future.

### IVa. Photoinjector Laser Temporal Shaping for Transverse Emittance Reduction

In our study, we note that Gaussian temporal intensity profiles are not ideal for photocathode drive lasers when aiming to minimize the transverse emittance of the electron beam. Transverse emittance refers to a measure of the spread of electron trajectories in a beam perpendicular to the direction of motion. Emittance quantifies the beam's quality, with lower emittance indicating a more focused beam, essential for high-precision applications. Emittance reduction refers to the process of narrowing the spread of particle positions and momenta, thereby improving beam focus and stability[65–67]. It has been shown that pulses with elliptical or flat-top intensity profiles in time can produce lower transverse emittance electron bunches[68]. For the 20-60 ps requirement of LCLS-II, temporal shaping is a non-trivial, long-standing challenge. Transform-limited picosecond pulses have too little spectral bandwidth to apply typical shaping schemes available to femtosecond optics and are too short to apply direct temporal amplitude shaping available to nanosecond pulses. To resolve this challenge, we have employed the dispersion-controlled nonlinear synthesis (DCNS) method described by Lemons et al.[59] to achieve temporal shaping in this regime. This method is a non-collinear sum frequency generation (SFG) scheme where the driving pulses are chirped with equal and opposite amounts of spectral phase and the sum-frequency pulse is generated with a temporal intensity profile roughly equivalent to the time-overlapped sum of the inputs.

For this effort, we use a Yb: KGW laser producing 256 fs pulses centered at 1024 nm with 40 W of power at configurable repetition rates from 100 kHz to 1 MHz. In order to generate the necessary 1024 nm (IR) inputs to achieve a 256 nm (UV) pulse with a full-width half maximum (FWHM) duration of 25 ps, we apply $\pm 2.561$ ps$^2$ of second order and $\pm 0.28$ ps$^3$ of third-order spectral phase to identical copies of our IR drive laser pulse via a matched stretcher/compressor pair[59]. The significant amount of second-order dispersion (SOD) requires a corresponding amount of the third-order dispersion (TOD) to maintain the ratio between TOD and SOD, which describe the general shape of a shaped pulse that is approximately invariant to the pulse duration. The IR pulses are employed as inputs in a noncollinear SFG scheme, utilizing type-I mixing within a 1 mm long BBO crystal. The crystal is oriented at an angle of 23.29 degrees, and the input beams are arranged internally at a half angle of 1.5 degrees for this process. Due to the non-collinear geometry of the harmonic generation, the angle at which the input beams intersect plays a crucial role in the conversion efficiency towards the targeted SFG beam, while also impacting the suppression of undesired parasitic second harmonic generation beams[69]. We demonstrated 30% conversion efficiency from the IR pulses to the SFG beam in initial experiments. This is commensurate with experimentally demonstrated conversion efficiencies for similar SFG schemes[59,68].

The spatial profile of the SFG pulse is shaped by the elliptical overlap of the two incident beams within the crystal, a result of the crossing angle. This profile is more representative of their combined overlapping geometry rather than just their individual spatial profiles. After transmitting through the second BBO crystal, Figure 7a and b display the achieved temporal intensity profile of the 256 nm beam measured with the 70 fs, 1030 nm oscillator pulses in an intensity cross-correlator. This profile is 26 ps FWHM and is characterized by a flattened intense region and faster rise/fall times compared to the Gaussian profile generated without phase addition. The 5 ps oscillations present on the plateau are likely the result of spectral filtering of the IR beams before mixing due to limited spectral acceptance of the dispersive elements. In our setup, the fundamental beam is spatially Gaussian with no ellipticity, SFG beam is produced with an ellipticity of 0.51, where the larger axis aligns parallel to the plane of the table and lies within the crossing plane. Despite this ellipticity, the beam profile maintains a smooth Gaussian shape in both directions. Figure 7c demonstrates the resulting profile of the SFG beam being used directly for further nonlinear conversion to 256 nm with an exacerbated ellipticity of 0.63, because of the crossing angle and phase matching in the crystals. This issue has been effectively addressed using a cylindrical lens telescope, employing a method akin to the correction of astigmatism in laser diodes. We have successfully implemented this technique in our setup. We focus the simulation on the photoinjector and the first 15 meters of acceleration (approximately 100 MeV). Within this constrained simulation range, we compare the simulated emittance on the cathode of a temporal Gaussian pulse to that of filtered DCNS UV pulses with three different spectral filters, as detailed in Figure 7d. This segment is crucial for our analysis because the evolution of the electron bunch within it is predominantly influenced by laser parameters, due to the internal space-charge forces not yet being mitigated by the effects of highly relativistic speeds. All spectral filters maintain at least 90% of the field power and the DCNS-shaped pulses demonstrate lower emittance after optimization of beam line settings such as magnets and phases. When utilizing a spectral filter with a bandwidth of 0.5 nm, longer bunch lengths exhibit better performance. Conversely, with a 1.0 nm filter, a higher density of data points is observed at shorter bunch lengths.

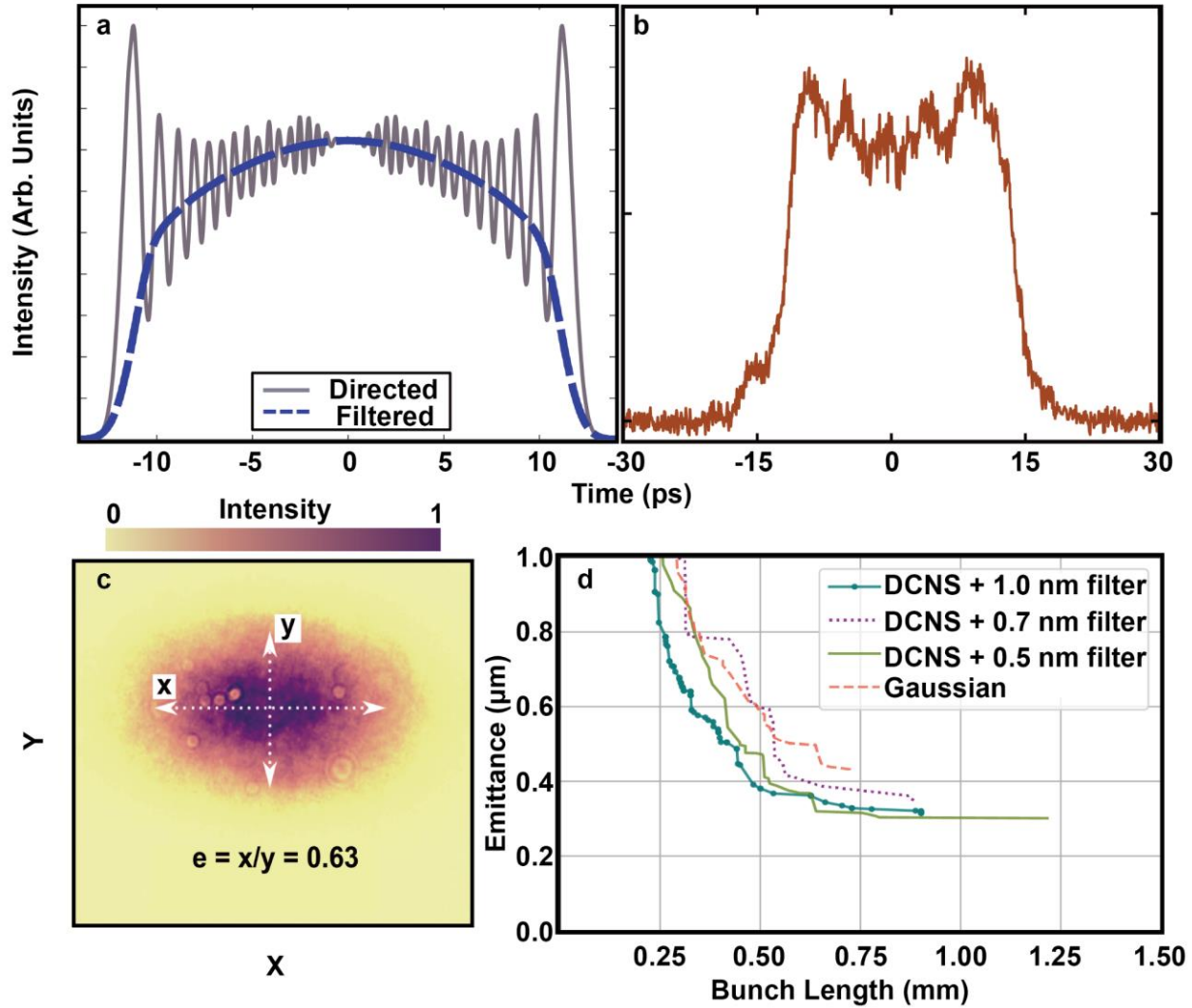

*Figure 7. (a) Numerically generated temporal profile of the sum-frequency pulse before applying a narrowband spectral filter (grey) and after (blue); (b) experimental temporal profile at 256 nm collected with cross-correlator with 70 fs, 1030 nm oscillator; (c) 256 nm spatial profile with an ellipticity of 0.63; (d) simulated emittance comparison between temporal Gaussian pulse and shaped pulses with 3 different spectral filters, where DCNS with a 0.5 nm spectral filter demonstrates improved emittance at all electron bunch lengths[59,70]. The charge used for optimization in part (d) is 100 pC. Copyright ©2022 American Physical Society.*

### IVb. Visible-range Responsive Photocathode Development

To reduce the normalized transverse emittance of the electron beam generated by the LCLS-II photoinjector, it is likely that SLAC will need to transition from UV-responsive photocathodes to using visibly responsive ones. There are two reasons for this. The first is that it is easier to generate the desired spatial and temporal drive laser shaping when working at the 2$^{nd}$ harmonic of the laser oscillator instead of the 3rd or 4th harmonic. This is important because the quality of the electron beam produced directly relates to the quality of the drive laser used. The second reason is that achieving a beam spot size of approximately 0.4 µm, with a charge of 100 pC) in a pulse duration of 30 ps, at an energy level of 100 MeV, will likely require a photocathode with an intrinsic emittance of less than 0.5 µm/mm.

Photocathode R&D has been carried out on various materials in recent years, and many have performed well in laboratory environments. However, only a small subset of them have ever been used in large-scale facilities (e.g., Cu, Cs-Te, and GaAs). None have demonstrated the ability to simultaneously generate 100 pC in a few tens of ps with < 0.5 µm/mm intrinsic emittance. Alkali-antimonide photocathodes (e.g. NaK2Sb or K2CsSb) are the best candidates for this. They have been used in several DC[43] and low-frequency normal-conducting radio frequency (NCRF) and superconducting radio frequency (SRF)[71,72] guns. Recent work suggests that they should be capable of 0.3 µm/mm intrinsic emittance with a quantum efficiency (QE) > 0.1% when illuminated at 690 nm[44] while meeting other operational requirements such as response time, lifetime, and stability at the high gradient and minimal dark current generation.

SLAC has started taking steps that will allow for the eventual operation of the LCLS-II photoinjector with visibly responsive photocathodes. However, the optimal choice of photocathode material and drive laser illuminating wavelength remains unknown. In order to achieve an emittance of less than 0.5 µm/mm, it is necessary for the photocathode to be illuminated with a laser wavelength that closely matches its work function. The work function is a fundamental property of the photocathode material, representing the minimum energy needed to remove an electron from its surface. When the laser wavelength is tuned near the work function, it optimizes the photoelectric effect, resulting in the efficient emission of electrons with minimal spread in their kinetic energies. This fine-tuning is crucial for producing an electron beam with low transverse emittance, thereby enhancing the beam's focus and quality for precise applications. Yet, this approach of using a laser wavelength that closely matches the work function, and hence results in low excess energy of the emitted electrons, will lead to a reduction in quantum efficiency (QE). If the QE is too low, the high drive laser power required to compensate for it may spoil the intrinsic emittance. For this and other practical reasons, the effort to adopt visibly photoresponsive photocathodes is moving forward cautiously, carefully evaluating candidates with lab-scale tests before demonstrating them in an operational environment.

Building on the discussion of optimizing photocathode performance in photoinjector systems, it's important to consider the role of high-power lasers, which are integral to these systems. Two popular gain materials for high-power lasers are titanium-doped sapphire (Ti:S) and ytterbium (Yb)-doped crystals. While Ti:S amplifiers are unrivaled in terms of the highest peak intensity with the shortest pulse durations, the power output of Ti:S lasers is generally lower compared to Yb lasers, which can operate at MHz repetition rates at several Watt-level peak powers, due to their low quantum defect and their ability to be directly pumped by high-power diodes. All these reduce thermal problems and improve conversion efficiency[73]. A particular advantage of Yb is that they can be used as a dopant in gain fibers, which further reduces the thermal load caused by optical pumping, as the heat is spread over a longer path and larger volume. Further, in a double-clad fiber, which has a small rare-earth doped single-mode core surrounded by a much larger multimode pump cladding, energy from a highly multimode low-brightness pump can be converted into a single-mode high-brightness laser beam guided through the single-mode core[73].

Regardless, as both Yb-doped and Ti:S lasers are limited in terms of wavelength, typically around 1.02-1.06 and 0.8 $\mu$m, respectively, their harmonics are not well-matched to the wavelength range demanded by visibly responsive photocathodes and thus require an optical parametric amplifier (OPA) to enable wavelength tuning. Most commercial OPAs use white-light-generated (WLG) pulses as a seed for amplification[74]. WLG produces a broadband pulse spanning the visible and near-infrared (NIR) by focusing an intense laser (pump) pulse inside a transparent medium[75]. However, an issue with WLG is low spectral energy density and high phase modulations in the spectral vicinity of the white light pump pulse where most of the energy is contained[76,77]. We are actively exploring an alternative approach by exploiting gain-managed nonlinear amplification in Yb-fiber amplifiers[77,78]. By sending in a parabolic pulse[77,79] through a double-clad Yb fiber amplifier, we can generate a coherent, smoothly broadband microjoule-level pulse spanning 1000-1200 nm that can be used as a seed for amplification in an OPA. For the seed preparation described here, an initial weak pulse at 1030 nm is first launched into a Yb-fiber amplifier chain that is pumped at 976 nm. The pulse propagation in an optical fiber amplifier is described by the non-linear Schrodinger equation (NLSE). In the high power limit, the asymptotic solution to the NLSE is a linearly chirped parabolic pulse that propagates self-similarly, i.e. it propagates while maintaining its parabolic pulse profile[80]. Self-similar pulses have a practical significance because they can be propagated at high power without pulse distortions resulting from optical wave-breaking[81]. However, we note that parabolic pulse amplification is not sufficient to obtain a seed pulse with high spectral energy density and a smooth broadened spectral profile suitable for tunable OPAs. Past the self-similar amplification limit, a new amplification regime emerges, known as gain-managed nonlinearity (GMN) [82]. In this regime, the pulse broadens in both the spectral and temporal domain and increases in energy, with the red-shifting gain and pulse spectra re-shaping each other in tandem. The red-shifting gain envelope allows the spectra to extend towards longer wavelengths, all the while increasing in pulse energy. Absorption near the blue end of the spectrum flattens the spectral profile near the initial launch pulse peak at 1030 nm. Because of the initial self-similar propagation, there is a range in the launched pulse characteristics to all evolve toward a similar output. When pumped by the second harmonic of a Yb-laser at 515 nm, we can generate signal wavelengths in the range of 1000-1200 nm. The second harmonic of this wavelength range (450-600 nm) directly corresponds to that required by visible-range responsive photocathodes. Owing to the high spectral density of fiber-amplifier parabolic pulses, about 20 times greater than WLG in the NIR, the OPA conversion efficiency of signal and idler combined can reach about 30% on average, based on simulations in Chi2D[77]. Combined with multiple OPA stages, such a system exploits the high average power scaling of Yb-based lasers and has the potential to generate picosecond visible pulses at high repetition rates suitable for photocathode excitation for state-of-the-art electron beamline facilities.

### IVc. Future applications

**Adaptive Spatio-Temporal Shaping and Beam Shaping for Attosecond X-ray Pulses**

Isolated attosecond X-ray pulses represent a remarkable achievement at LCLS, marking a significant advancement in ultrafast X-ray science. This capability is part of a broader initiative known as x-ray laser-enhanced attosecond pulse generation (XLEAP). Notably, attosecond XFELs have pushed the boundaries further, demonstrating the generation of pulses with more than six orders of magnitude higher pulse energy when compared to conventional table-top high-harmonic generation (HHG) sources, showcasing the remarkable progress in XFEL technology[1]. Within the context of the LCLS-II project, there is a dedicated focus on developing cutting-edge laser shaping techniques to harness the full potential of these attosecond X-ray pulses. This involves two primary methods: laser heater shaping and photocathode shaping. Laser heater shaping is focused on manipulating the longitudinal distribution of the electron beam, thereby controlling its energy spread and minimizing the effects of microbunching instabilities. On the other hand, photocathode shaping involves the precise modulation of the spatial and temporal profiles of the laser pulses used to emit electrons from the photocathode. This precision shaping of the electron emission pattern is crucial for optimizing the beam quality and reducing transverse emittance, thereby enhancing the coherence and intensity of the resultant X-ray pulses. These innovative techniques enable the precise modulation of the electron beam's current profile, effectively perturbing it to create a high-current spike that drives the attosecond emission. Additionally, they play a pivotal role in seeding the microbunching instability, which is essential for achieving the desired attosecond pulse characteristics. The combination of these shaping methodologies not only represents a technical breakthrough but also opens new horizons in the realm of attosecond science and its applications[55,83].

### Laser Heater Beam Shaping for Attosecond Emission

Localized heating of the electron beam via a short Gaussian laser heater pulse located after the photoinjector is used to introduce energy spread in the electron beam. The energy spread is compressed into a high-current spike. Changing the characteristics of the laser heater pulse manipulates the current spike produced. The short-duration current spike may then be used to lase on and generate attosecond X-ray pulses with the chirp-taper method[84]. We selectively heat part of the electron beam, the one generated by the photoinjector, by stacking pairs of laser heater pulses. These pulses in the laser heater are strategically timed and overlapped to achieve the desired modulation in the electron beam's energy spread. As the XFEL is sensitive to energy spread, we control the output FEL duration to be on the femtosecond scale by varying the delay between laser heater pulses. We implement a wider laser heater pulse, which introduces a small modulation following the initial bunch compressor. This small modulation is enough to generate a current spike near the end of the bunch after the large compression factor in the second bunch compressor chicane and at the entrance of the soft X-ray undulator.

### Photoinjector Laser Beam Shaping for Attosecond Emission

A density perturbation is introduced to the electron beam at the photoinjector cathode by adding a small temporal perturbation on top of the production cathode laser pulse. This can be achieved by either shaping the photoinjector laser pulse so it has a sharp (~ 1-2 ps) perturbation on it, or by superimposing a short laser pulse on top of the standard photoinjector pulse. Alternatively, the same effect can be achieved using a pulse stacker and generating two slightly separated Gaussian pulses, resulting in a power profile with a modulation. The resulting current modulation is amplified by the microbunching instability and generates a high-current spike at the entrance of the undulator. Similar to the laser heater shaping method, this technique allows for the implementation of an arbitrary pulse train structure. This is achieved by finely tuning the temporal profile of the photoinjector laser pulses, either through the direct shaping of individual pulses or by stacking and manipulating multiple pulses. The flexibility of this approach enables us to selectively shape the electron beam for specific LCLS-II undulator beamlines. The LCLS-II system includes multiple undulator beamlines, each capable of independent operation. By applying this technique, we can tailor the beam characteristics for one undulator beamline, such as optimizing for attosecond pulse generation, without disrupting the standard operation of the other beamlines. This selective shaping provides versatility in conducting diverse experiments simultaneously, leveraging different beam properties as required by various research objectives.

### Next Generation Real-time Adaptive Photoinjector Shaping

Beyond manipulating the electron beam after its generation, efforts are underway for UV pulse shaping by temporal shaping the pre-upconversion IR pulse. Specifically, the desired shaping apparatus will utilize a programmable acousto-optic dispersive filter to execute pre-amplification shaping of the Infrared (IR) pulse, facilitating high-rate amplitude and phase modulation[85]. This translates to the shaping of the upconverted UV pulses for interaction with the photocathode. However, the nonlinearity in amplification and upconversion complicate the programmable shaping, necessitating a machine learning (ML) approach to learn the desired shaping parameters to achieve the finalized UV beam.

In order to generate the required amount of data for these ML studies, we have developed a start-to-end software model of the photoinjector laser to explore the shaping parameter space[86,87]. Our current studies focus on tuning the software models to closely match the experimental system and developing ML-enhanced simulation techniques to speed up the data generation process[88]. Using these methods, we can then generate the proper datasets for training the ML networks for learning how to control the pulse shaper to reach desired UV shapes while simultaneously feeding back experimental results to improve the software models[86]. This adaptive shaping and complementary photoinjector laser shaping enable the manipulation of the electron beam at the photocathode to generate a specific temporal profile and, ultimately, an attosecond XFEL pulse.

### V. CONCLUSIONS

In conclusion, we provide a detailed overview of the most critical photoinjector infrastructure components for the LCLS-II and the future LCLS-II-HE facilities. We have discussed the different subsystems, including the photocathode laser system, laser heater laser, and beam transport, as well as the engineering challenges that are currently being addressed. The ongoing R&D efforts for photoinjector laser pulse temporal and spatial shaping, visible-range photocathode development, and computationally intelligent adaptive spatiotemporal shaping were also highlighted. While this paper addresses the primary, pivotal challenges of the LCLS-II facility, it's acknowledged that specific issues, such as UV beam quality, remain. Ongoing efforts, including exploring Four-Wave Mixing (FWM) architectures[87,89-90] and the potential adoption of green beam irradiated photocathodes, aim to refine these aspects further. The development of the LCLS-II photoinjector not only contributes to advancing X-ray science but also forges new exploratory pathways across a broad scientific spectrum. Its ability to generate high-quality electron beams catalyzes the creation of brighter, faster, and more coherent X-ray pulses, allowing for high resolution and unprecedented investigative precision in spatial and temporal regimes. This enhanced capacity will enable the exploration of atomic and molecular structures more deeply, comprehend dynamic biological processes more accurately, and explore material properties under extreme conditions more effectively. It promises to drive innovation by inspiring novel experimental methodologies and advancing technological frontiers, thereby reshaping our understanding of the world at the most fundamental levels.

## Acknowledgments

The authors would like to thank the support from SLAC National Accelerator Laboratory, the U.S. Department of Energy (DOE), the Office of Science, Office of Basic Energy Sciences under Contract No. DE-AC02-76SF00515, No. DE-SC0022559, No. DE-SC0022464, No. DE-FOA-0002859, the National Science Foundation under Contract No. 2231334, and the U.S. Department of Defense under a National Defense Science and Engineering Fellowship. Additionally, we thank the useful discussion with Zhirong Huang about laser heater shaping.

## Author Contributions

S.G. and A.M. played a pivotal role in the development of the LCLS-II photoinjector laser system. H.Z. and S.C. conceptualized the original manuscript. All authors were involved in the overall development of the LCLS-II program and contributed to the manuscript.